\documentclass[preprints,review,accept,pdftex,moreauthors]
{Definitions/mdpi}

\pdfoutput=1 

\usepackage{lineno}
\usepackage{pifont}
\usepackage{array}
\Title{Stellar microlensing surveys as a probe of Primordial Black Holes: status and prospects}

\TitleCitation{Title}

\Author{Anne M. Green  $^{1}$\orcidA{} }

\AuthorNames{Anne Green }

\isAPAStyle{%
       \AuthorCitation{Lastname, F., Lastname, F., \& Lastname, F.}
         }{%
        \isChicagoStyle{%
        \AuthorCitation{Lastname, Firstname, Firstname Lastname, and Firstname Lastname.}
        }{
        \AuthorCitation{Lastname, F.; Lastname, F.; Lastname, F.}
        }
}

\address{%
$^{1}$ \quad School of Physics and Astronomy, University of Nottingham, University Park, Nottingham, NG7 2RD, UK; anne.green@nottingham.ac.uk
}

\corres{Correspondence: anne.green@nottingham.ac.uk}

\abstract{Stellar microlensing surveys are a powerful tool for probing dark matter in the form of planetary and stellar mass compact objects (COs), in particular primordial black holes (PBHs). Under standard assumptions, current observations exclude COs in the mass range 
$10^{-11} \lesssim M/M_{\odot} \lesssim 10^{4}$
making up all of the dark matter.  
We provide an overview, aimed at theorists working on PBHs, of the history, theory, observational status, and future prospects of the field.}


\begin{document}


\section{Introduction and history}

Stellar microlensing is the temporary magnification of a star which occurs when a compact object crosses the line of sight. It was first studied by Einstein, who concluded that 'there is no great chance of observing this phenomenon' \cite{Einstein:1936llh}. However in 1986 Paczy\'{n}ski showed that by monitoring millions of stars in the Magellanic Clouds (MC) it would be possible to probe compact objects (COs) in the Milky Way (MW) halo \cite{Paczynski:1985jf}.

Various collaborations started microlensing surveys in the early 1990s, and the first MC candidate events were published in 1993 by EROS \cite{1993Natur.365..623A}, and MACHO \cite{SupernovaCosmologyProject:1993faz}. By the late 1990s the MACHO collaboration had observed significantly more events than expected from stars in the MC or MW disk. This indicated that a significant fraction of the Milky Way halo could be in the form of roughly Solar mass COs \cite{MACHO:1996qam,MACHO:2000qbb}. Baryon budget arguments excluded such a large population of astrophysical objects (e.g. white dwarfs) \cite{Fields:1999ar}. Therefore it was suggested, e.g. Refs.~\cite{Yokoyama:1995ex,Jedamzik:1998hc}, that these events could be due to Primordial Black Holes (PBHs), black holes that may form in the early Universe from large overdensities \cite{Zeldovich:1967lct,Hawking:1971ei}. However the EROS \cite{1996A&A...314...94A,2000A&A...355L..39L,2007A&A...469..387T} and OGLE \cite{Wyrzykowski:2009ep,Wyrzykowski:2011tr} collaborations did not see a significant excess of events, and instead constrained planetary and stellar mass COs to make up less than of order 10\% of the MW halo.

In 2016 the discovery of gravitational waves from mergers of multi-Solar mass PBHs by LIGO-Virgo \cite{LIGOScientific:2016aoc} generated a surge of interest in PBH dark matter (DM) \cite{Bird:2016dcv,Clesse:2016vqa,Sasaki:2016jop,Carr:2016drx}. However microlensing observations over two decades by the OGLE collaboration have failed to find the large, $\gtrsim 10^{2}$, number of events expected (under standard assumptions) if a significant fraction of the MW halo is in the form of stellar mass COs \cite{Mroz:2024wag}. These long duration observations have also increased the largest CO mass probed to $\sim 10^{4} M_{\odot}$, while high cadence observations by Niikura et al. have decreased the lowest mass to $\sim 10^{-11} M_{\odot}$ \cite{Niikura:2017zjd}.

We overview how stellar microlensing surveys probe the fraction of COs, such as PBHs \footnote{Throughout, we use CO when covering material that applies to compact objects in general, and PBH for material that is specific to Primordial Black Holes.}, in the MW halo by observing stars in the MC or M31  \footnote{  We do not cover constraints on CO DM \cite{Diego:2017drh,Muller:2024pwn} from supermagnified stars (a star in a galaxy that is magnified by a large factor by lensing by a galaxy cluster, with the star then magnified further by microlensing by a CO) \cite{Venumadhav:2017pps}.}. Our aim is to provide an introduction to this field that is accessible to theorists who are interested in PBHs, and specifically the possibility that planetary and stellar mass PBHs make up a significant fraction of the DM. First, in Sec.~\ref{sec:theory}, we review the theory of stellar microlensing, before focusing on observations in Sec.~\ref{sec:obs}. We then look at future prospects in Sec.~\ref{sec:future}, before concluding with a brief summary in Sec.~\ref{sec:summary}.

For an overview of the history of PBHs, see Ref.~\cite{Carr:2024nlv}.  Refs.~\cite{Green:2020jor,Carr:2026hot} review the formation of PBHs and the full range of observational probes of their abundance. 
For a review of the theory of microlensing see, e.g., Ref.~\cite{2012RAA....12..947M}, while Refs.~\cite{Moniez:2010zt,Jetzer:2014uca} overview the observational status as of the 2010s.

\section{Theory}
\label{sec:theory}

In this Section we overview the theory of stellar microlensing. First, in Sec.~\ref{subsec:basics}, we cover the basics of photometric microlensing, including 
individual events (Sec.~\ref{subsubsec:indivdual events})
and the optical depth and event rate (Sec.~\ref{subsubsec:opticaldepthrate}). We then look at
the differential event rate and parameter constraints (Sec.~\ref{subsec:diffeventrate}) and non-standard events (Sec.~\ref{subsec:nonstandard}).
Finally we conclude, in Sec.~\ref{subsec:astrometric}, with a brief overview of astrometric microlensing.

\subsection{Basics}
\label{subsec:basics}
\subsubsection{Individual events}
\label{subsubsec:indivdual events}
In this subsection we overview the key properties of individual 'standard' photometric microlensing events where a point source is lensed by a point lens. For more details see, e.g., the original microlensing paper by Paczy\'{n}ski \cite{Paczynski:1985jf} or the review by Mao \cite{2012RAA....12..947M}.

Microlensing occurs when the splitting of images caused by gravitational lensing is too small to be resolved, and instead the source is amplified by a factor \cite{Refsdal:1964yk,Paczynski:1985jf}:
\begin{equation}
\label{eq:amp}
A = \frac{u^2 + 2}{u \sqrt{u^2 + 4}} \,,
\end{equation}
where $u=b/R_{E}$, $b$ is the distance of the lens from the line of sight to the source and
$R_{\rm E}$ is the Einstein radius
\begin{equation}
R_{{\rm E}} = 2 \left[ \frac{ G M D_{\rm L} (D_{\rm S} - D_{\rm L})}{c^2 D_{\rm S} } \right]^{1/2} \,, \label{re}
\end{equation}
where $G$ is the Gravitational constant, $M$ is the mass of the CO, $D_{\rm S}$ is the distance from the observer to the source , and $D_{\rm L}$ is the distance to the lens.  For a Large Magellanic Cloud (LMC) microlensing survey, $D_{\rm S} = 49.6 \, {\rm kpc}$ \cite{2019Natur.567..200P} and
\begin{equation}
R_{\rm E} \sim 10^{-4} \, {\rm pc} \sqrt{x(1-x)} \left( \frac{M}{M_{\odot}} \right)^{1/2} \left( \frac{D_{\rm s}}{50 \, {\rm kpc}} \right) \,,
\end{equation}
where $x=D_{\rm L}/D_{\rm S}$ is the fractional distance of the lens along the line of sight ($ 0 \leq x \leq 1$).

A microlensing event is usually defined to occur when $u \leq 1$, which corresponds to $A \geq A_{\rm T} = 1.34$. The timescale of the event, $\hat{t}$, is defined as the time taken to cross the Einstein diameter or radius. The EROS and OGLE collaborations use the Einstein radius crossing time, while the MACHO collaboration and this review use the Einstein diameter crossing time, $\hat{t} = 2 R_{\rm E}/v_{\rm t}$, where $v_{\rm t}$ is the transverse speed of the lens relative to the line of sight. For lensing of stars in the LMC by CO in the MW halo typically
\begin{equation}
\hat{t} \sim 1 \, {\rm yr} \, \sqrt{x (1-x)} \left( \frac{M}{M_{\odot}}  \right)^{1/2}  \left( \frac{v_{\rm t}}{ 200 \, {\rm km \, s}^{-1}} \right)^{-1} \,.
\end{equation}
Due to the mass dependence of the timescale, observations over a long duration are needed to probe large masses, while high cadence observations (i.e. at short intervals) are required to probe small masses.

Because the timescale, $\hat{t}$, depends on the lens mass, distance and velocity, it is not possible to measure the lens mass from the timescale of a single standard micolensing event (this is often referred to as the microlensing degeneracy). However, as we will see in Sec.~\ref{subsec:diffeventrate}, the lens mass can be probed statistically from the distribution of timescales, given a model for the lens density and velocity distributions.
The Einstein crossing time, $\hat{t}$, is different from the time for which the amplification exceeds the threshold value $A_{\rm T} = 1.34$, $t_{\rm e}$, and has observational and theoretical advantages over $t_{\rm e}$ \footnote{These advantages are well known in the observational microlensing community, but are hard to find original sources for.}. Observationally,  determining when the amplification crosses a particular threshold (and hence $t_{\rm e}$) is challenging, whereas $\hat{t}$ is found by fitting to the entire light curve.
Theoretically, $t_{\rm e}$ depends on the impact parameter (as well as the lens mass, distance and velocity), which makes the calculation of the differential event rate more complicated.

\begin{figure}[t]
\isPreprints{\centering}{} 
\includegraphics[width=7.5 cm]{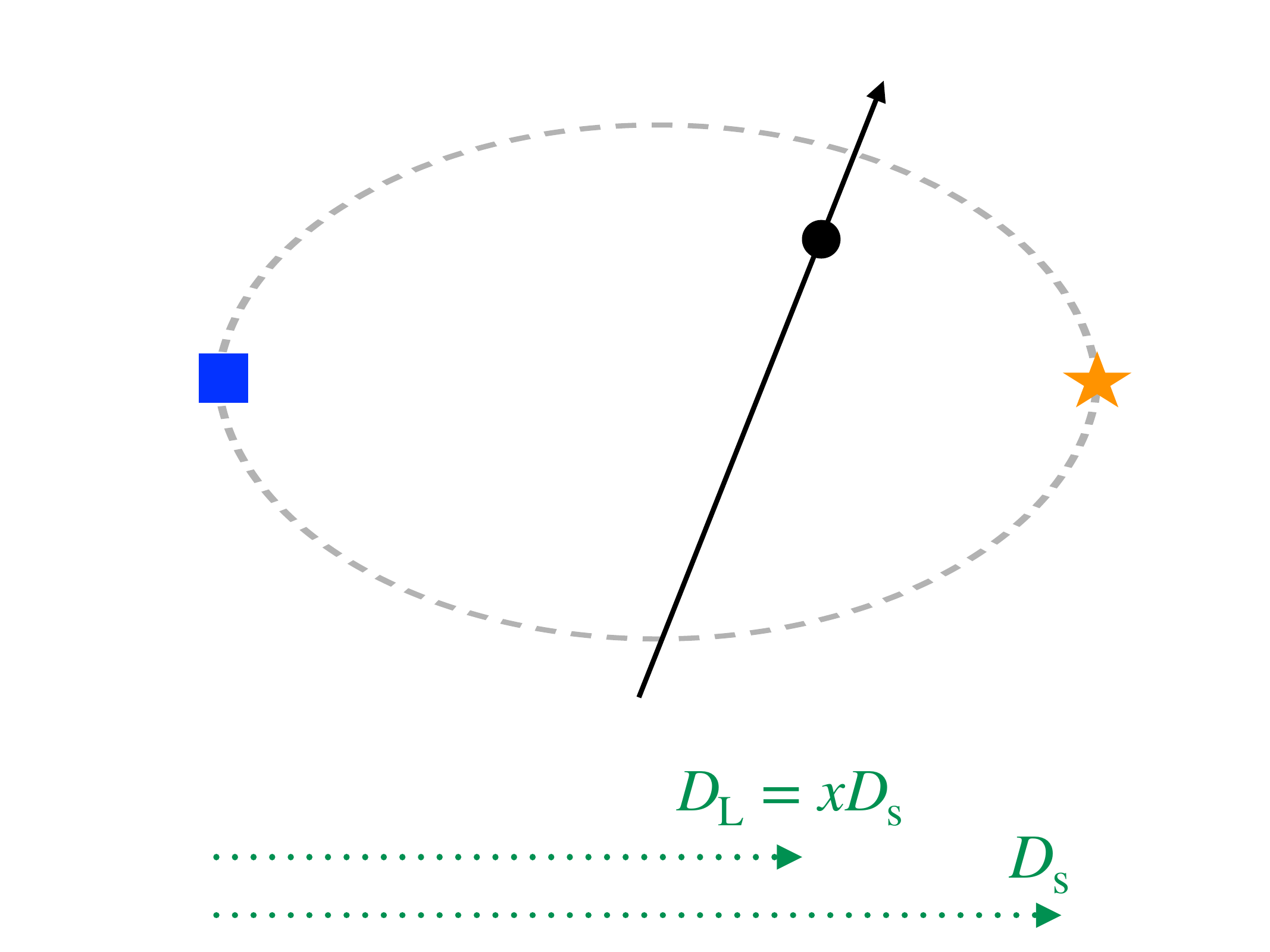}
\includegraphics[width=7.5 cm]{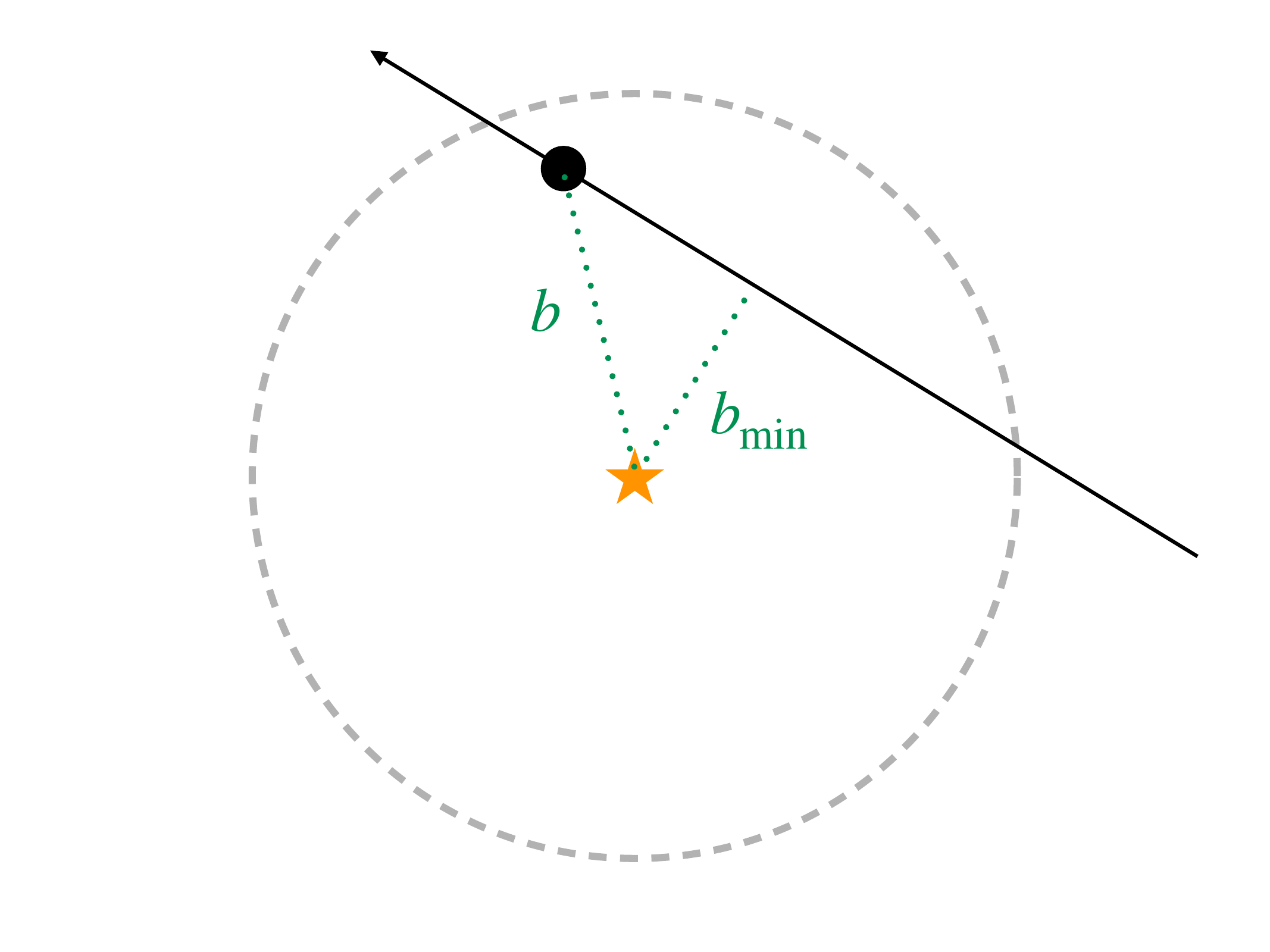}
\caption{The geometry of a stellar microlensing event. The left and right panels show the view perpendicular to and along the line of sight respectively. The solid black and dashed grey lines show the trajectory of the compact object and the Einstein radius, $R_{\rm E}$, respectively. The orange star, black circle and blue square denote the source, the lens and the observer respectively. In the left panel the dotted green lines show the distance from the observer to the lens and source, $D_{\rm L}= x D_{\rm S}$ and $D_{\rm S}$ respectively. In the right panel the green dotted lines show the impact parameter, $b$, and its minimum value $b_{\rm min}$.
\label{fig:geom}}
\end{figure}

\begin{figure}[t]
\isPreprints{\centering}{} 
\includegraphics[width=12 cm]{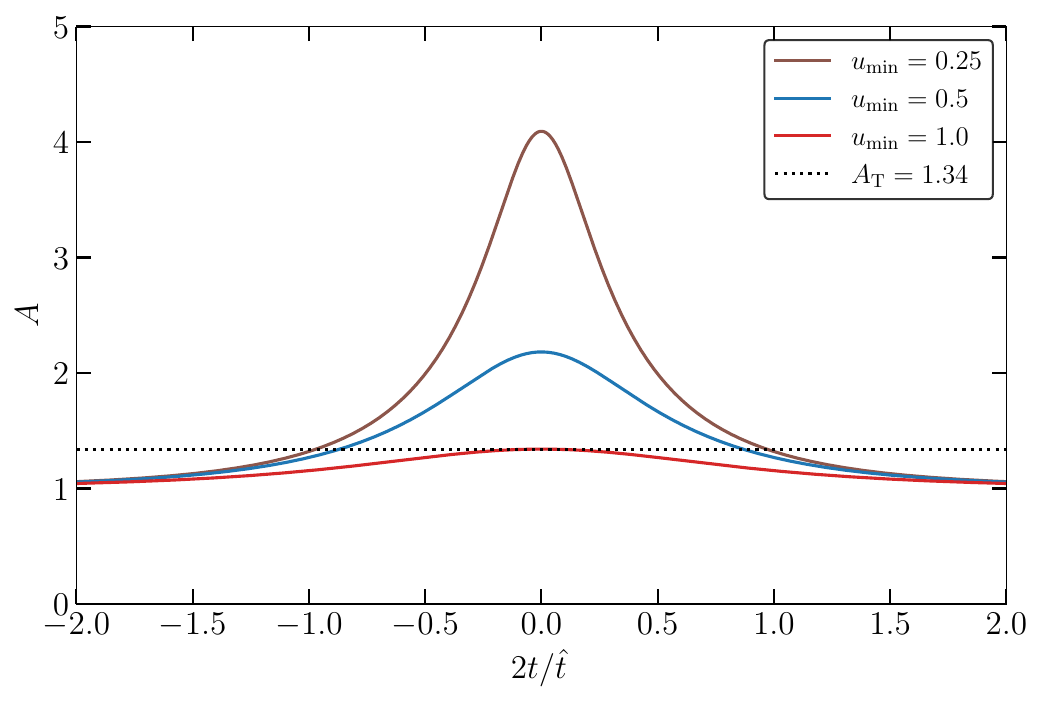}
\caption{The magnification factor, $A$, Eq.~(\ref{eq:ut}),
as a function of $2 t/\hat{t}$ for $u_{\rm min}=b_{\rm min}/R_{\rm E} = 0.25, 0.5$ and $1.0$ (from top to bottom). The horizontal dotted black line shows the threshold for a microlensing event, $A_{\rm T} =  1.34$. 
\label{fig:lightcurve}}
\end{figure}

The impact parameter relative to the Einstein radius, $u= b/R_{\rm E}$, can be written as
\begin{equation}
\label{eq:ut}
u= \left( \frac{2 t^2}{\hat{t}^2}  + u_{\rm min}^2 \right)^{1/2} \,,
\end{equation}
where $u_{\rm min}= b_{\rm min}/R_{\rm E}$, $b_{\rm min}$ is the minimum value of the distance of the lens from the line of sight to the source, and time $t$ is defined so that $t=0$ when $b=b_{\rm min}$. Figure \ref{fig:geom} shows the geometry of a microlensing event, as viewed perpendicular to and along the line of sight, including $R_{\rm E}$, $D_{\rm S}$, $D_{\rm L}$, $b$ and $b_{\rm min}$. Figure \ref{fig:lightcurve} shows the magnification factor $A$, given by Eq.~(\ref{eq:ut}),  as a function of $2 t/\hat{t}$ for different values of $u_{\rm min}$. This light curve, which is symmetric and achromatic (i.e.~the same for all wavelengths), is often referred to as the Paczy\'{n}ski light curve.

\subsubsection{Optical depth and event rate}
\label{subsubsec:opticaldepthrate}
The optical depth, $\tau$, is the probability that a source star is being lensed at a given instant, i.e. that it falls within the Einstein radius of a lens: 
\begin{equation}
\tau = \int_{0}^{D_{\rm S}} n(D_{\rm L}) \pi R_{\rm E}^2(D_{\rm L}) \, {\rm d}D_{\rm L} = 4 \pi G \int_{0}^{D_{\rm S}} \rho(D_{\rm L}) \frac{D_{\rm L}(D_{\rm S}-D_{\rm L})}{D_{\rm S}}  \, {\rm d}D_{\rm L}
\,,
\end{equation}
where $n(D_{\rm L})$ and $\rho(D_{\rm L})=n(D_{\rm L})/M $ are the number density and (mass) density of lenses respectively.
Note that the dependence on the lens mass, $M$, of the number density and the square of the Einstein radius cancel, and hence the optical depth depends on the total mass density of lenses along the line of sight, but not their individual masses. For a very simple model of the Milky Way (MW) halo with constant density, and a circular speed $v_{\rm c} = \sqrt{G M(<D_{\rm S})/D_{\rm S}} \approx 220 \, {\rm km \, s}^{-1}$, $\tau = v_{\rm c}^2/2c^2 \sim 10^{-6}$ \cite{2012RAA....12..947M}. This small optical depth means that to detect microlensing events it is necessary to monitor a large number, i.e. millions, of stars. Furthermore, it is extremely unlikely that a single star will undergo two independent microlensing events.

A more useful quantity is the  rate, ${\Gamma}$, at which events occur, and we present a modified version of  the illustrative calculation outlined in Ref.~\cite{2012RAA....12..947M}. The area, ${\rm d} \tilde{A}$, swept out by a lens in a time interval, ${\rm d} t$, is given by the product of the diameter of the Einstein ring, $2 R_{\rm E}$, and the distance traveled, $v_{\rm t} {\rm d} t$: ${\rm d} \tilde{A}= 2 R_{\rm E} v_{\rm t} {\rm d} t = 4 R_{\rm E}^2 {\rm d} t/ \hat{t}$. The probability, ${\rm d} \tau$, of an individual star undergoing a new microlensing event is given by
\begin{equation}
{\rm d} \tau =  n(D_{\rm L})\, \pi R_{\rm E}^2 \, {\rm d} D_{\rm L}  \,.
\end{equation}
If $N_{\star}$ is the number of stars being monitored, then the number of new microlensing events in time ${\rm d} t$, ${\rm d} N$, is given by
\begin{equation}
{\rm d} N   = N_{\star} \int_{0}^{D_{\rm S}} n(D_{\rm d}) {\rm d} \tilde{A} \, {\rm d} D_{\rm d} =  N_{\star} \int_{0}^{D_{\rm S}} n(D_{\rm d}) \frac{4 R_{\rm E}^2}{\hat{t} } \, {\rm d} t \, {\rm d} D_{\rm d} = \frac{4 N_{\star}}{\pi} \int_{0}^{D_{\rm S}} \frac{1}{\hat{t}} {\rm d} \tau \, {\rm d} t \,,
\end{equation}
and hence the rate at which new events occur, $\Gamma$, is given by
\begin{equation}
\Gamma \equiv \frac{ {\rm d} N}{{\rm d} t} = \frac{4 N_{\star}}{\pi} \int_{0}^{D_{\rm S}} \frac{{\rm d} \tau}{\hat{t}} \,.
\end{equation}

Assuming that all events have the same timescale $\hat{t}$, leads to a rough relationship between the event rate and optical depth:
\begin{equation}
\Gamma \sim \frac{4 N_{\star}}{\pi \hat{t}} \tau \,.
\end{equation}

\subsection{Differential event rate and parameter constraints}
\label{subsec:diffeventrate}
Since the timescale of a microlensing event depends on the CO mass (as well as its position and velocity), the differential event rate, ${\rm d} \Gamma/ {\rm d} \hat{t}$, i.e. the rate at which events of timescale $\hat{t}$, occur can be used to probe the CO mass. The standard expression for the LMC differential event rate, for a halo composed entirely of COs ($f=1$, where $f$ is the fraction of the total mass of the halo in CO) with mass $M$ and a smooth density distribution $\rho(r)$ is~\cite{MACHO:1996qam}:  
\begin{equation}
\label{df}
\frac{{\rm d} \Gamma_{\rm MW}}{{\rm d} \hat{t}} =  \frac{32 D_{\rm S}}
                 { M {\hat{t}}^4
              v_{{\rm c}}^2}
              \int^{1}_{0} \rho(x) R^{4}_{{\rm E}}(x)
              \exp{\left[-Q(x)\right]}  {\rm d} x \,, 
\end{equation}
where  $Q(x)= 4 R^{2}_{{\rm E}}(x) / (\hat{t}^{2} v_{{\rm c}}^2)$. Note that Eq.~(\ref{df}) assumes that the velocity distribution, $f({\bf v})$, is Maxwellian 
\begin{equation}
\label{maxwellian}
f({\bf v}) \, {\rm d}^3 {\bf v} = \frac{1}{\pi^{3/2} v_{\rm c}^{3}} \exp{ \left( -\frac{v^2}{v_{\rm c}^2} \right)} \, {\rm d}^{3} {\bf v} \,, 
\end{equation}
however this is only true for an isotropic, isothermal sphere.

\begin{figure}[t]
\isPreprints{\centering}{} 
\includegraphics[width=12 cm]{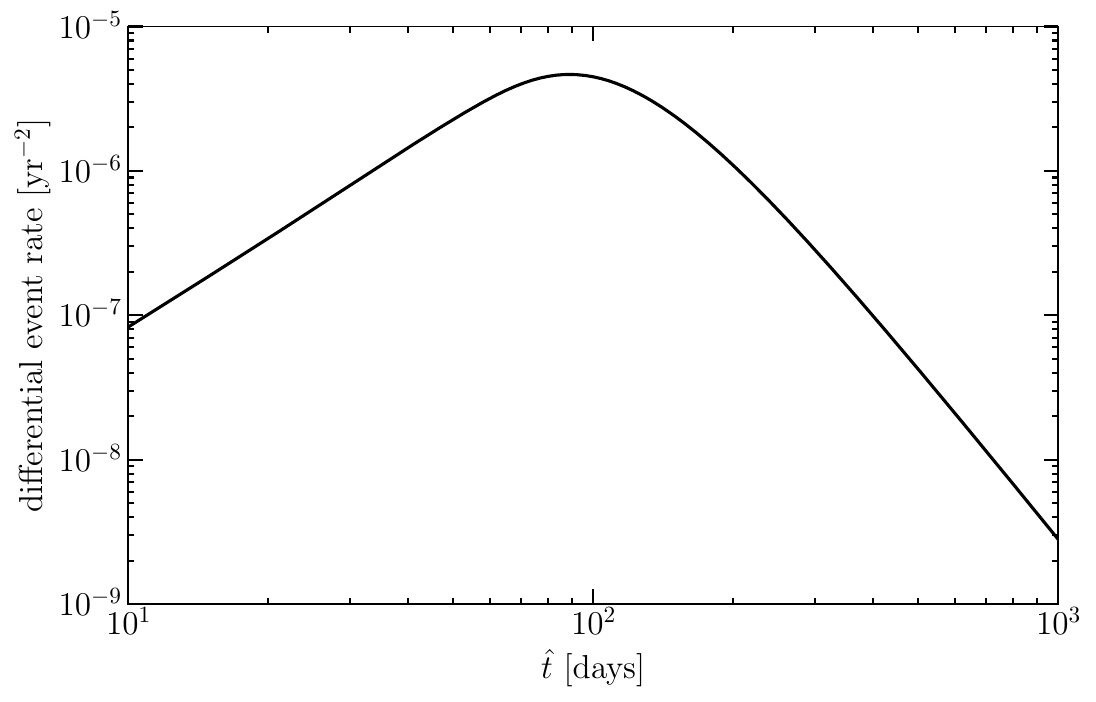}
\caption{The differential event rate, ${\rm d} \Gamma / {\rm d} \hat{t}$, as a function of Einstein diameter crossing time, $\hat{t}$, for a standard halo (see text for details) composed entirely, $f=1$, of COs with mass $M=1 M_{\odot}$ as given by Eq.~(\ref{dfsh}).
\label{fig-diffeventrate}}
\end{figure}   

The standard halo model historically assumed in microlensing studies (`Model S' in e.g. Ref.~\cite{MACHO:1996qam}) has a density profile
\begin{equation}
\rho(r) = \rho_{0} \, \frac{r_{{\rm c}}^2 + r_{0}^2}{r_{{\rm c}}^2 + r^2} \,,
\label{rhor}
\end{equation}
and local dark matter density $\rho_{0} \equiv \rho(r_{0}) = 0.0079 M_{\odot} {\rm pc}^{-3}$, core radius $r_{{\rm c}} = 5$ kpc, 
Solar radius $r_{0} = 8.5$ kpc and circular speed $v_{\rm c}  =220 \, {\rm km \, s}^{-1}$ \footnote{There have been changes to some of best fit parameter values,
for instance the Solar radius has been measured as $r_{0} = (8.18 \pm 0.01\pm 0.02)$ kpc by the GRAVITY collaboration \cite{2019A&A...625L..10G}. However, these changes have a smaller effect on the differential event rate than uncertainties in the density profile.}. 
 For the standard halo model, the differential event rate, Eq.~(\ref{df}), becomes~\cite{MACHO:1996qam} 
\begin{equation}
\label{dfsh}
\frac{{\rm d} \Gamma_{\rm MW}}{{\rm d} \hat{t}} = \frac{512 \rho_{0} 
          (R_{{\rm c}}^2 + R_{0}^2) D_{\rm S} G^2  M }
             {  {\hat{t}}^4 {v_{{\rm c}}}^2 c^4}
              \int^{1}_{0} \frac{x^2 (1-x)^2}{A + B x + x^2}
            \exp{\left[ -Q(x)\right] }{\rm d} x \,, 
\end{equation}
where $A=(r^2_{{\rm c}}+ r^2_{0})/D_{\rm S}^2$, $B=-2(r_{0}/D_{\rm S}) \cos{b}
\cos{l}$ and $b = -32.8^{\circ}$ and $l = 281^{\circ}$ are the Galactic
latitude and longitude, respectively, of the LMC.
Figure \ref{fig-diffeventrate} shows the differential event rate for a standard halo composed entirely, $f=1$, of COs with $M=1 M_{\odot}$. For different masses,
the differential event rate scales as $M^{-1/2}$ and the timescales as $M^{1/2}$ \cite{Griest:1990vu}. For short-timescale events ${\rm d} \Gamma / {\rm d} \hat{t} \propto \hat{t}^{2}$ while for long-timescale events ${\rm d} \Gamma / {\rm d} \hat{t} \propto \hat{t}^{-4}$, independent of the CO distribution \cite{MACHO:1995udp,Zhao:1995qi,Mao:1996hr}. 

There are also contributions to the total differential event rate, ${\rm d} \Gamma/{\rm d} \hat{t}$, from stars ('$\star$') in the MW disc and LMC, and COs in the LMC:
\begin{equation}
\label{totdifrate}
\frac{{\rm d} \Gamma}{{\rm d} \hat{t}}
= f \frac{{\rm d} \Gamma_{\rm MW}}{{\rm d} \hat{t}} + \frac{{\rm d} \Gamma_{{\rm MW}, \star}}{{\rm d} \hat{t}} + \frac{{\rm d} \Gamma_{{\rm LMC}, \star}}{{\rm d} \hat{t}} + f \frac{{\rm d} \Gamma_{\rm LMC}}{{\rm d} \hat{t}} \,.
\end{equation}
Lensing by stars in the LMC, a possibility that was pointed out by Sahu  \cite{Sahu:1994pj}, is often referred to as 'self-lensing'.
The self-lensing rate is largest for observing fields in the centre of the LMC, where the number of stars is highest. The MACHO collaboration found, using models of the LMC disc from Ref.~\cite{Gyuk:1999sb}, an optical depth $\tau_{{\rm LMC}, \star} \sim (1-3) \times 10^{-8}$.
Using the same LMC model with some minor modifications, OGLE III and IV (which has a wider observing field) found 5.7 events expected from stars in the LMC (compared with the $10^{2-3}$ expected if all of the halo is in planetary or stellar mass COs) \cite{Mroz:2024mse}.
The MACHO collaboration found an optical depth for microlensing by stars in the MW disc, $\tau_{{\rm MW}, \star} \lesssim 10^{-8}$, lower than their estimate for self-lensing \cite{MACHO:2000qbb}. However more recent studies have found significantly larger MW disc rates, comparable with, or even larger, than the self-lensing rate, in part due to a larger thick disc contribution \cite{2011MNRAS.416.1292C,Mroz:2024mse}.  For the two MW models they considered \cite{Han:2003ws,Cautun:2019eaf}, OGLE III and IV found 7.0 and 14.7 expected events respectively from stars in the MW disc  \cite{Mroz:2024mse}. Note, however, that the relative numbers of events from different lens populations depend on the fields observed, and the detection efficiency (see below).

The total expected number of events, $N_{\rm exp}$, is given by
\begin{equation}
\label{eq:nexp}
N_{\rm exp} = E \int_{0}^{\infty} \frac{{\rm d} \Gamma}{{\rm d} \hat{t}} \epsilon (\hat{t}) \, {\rm d} \hat{t} \,,
\end{equation}
where $E$ is the exposure (which is equal to the number of stars observed, $N_{\star}$, multiplied by the duration of the survey), $\epsilon (\hat{t})$ is the efficiency, i.e. the probability that a microlensing event with timescale $\hat{t}$ is detected (see Sec.~\ref{subsec:gen}), and ${\rm d} \Gamma/{\rm d} \hat{t}$ is the total differential event rate given by Eq.~(\ref{totdifrate}). 
If no events are detected, then the $95\%$ confidence limit on the fraction of the MW halo in COs can be calculated as $f=3.0/N_{\rm exp}(f=1)$, where $N_{\rm exp}(f=1)$ is the expected number of events if $f=1$.
Surveys with a large exposure expect to observe events due to stars in the MW disc and/or LMC, and in this case constraints on the CO mass and halo fraction are calculated using the likelihood \cite{MACHO:2000qbb,Mroz:2024mse}: 
\begin{equation}
\label{eq:like}
{\cal L}(f, M) = \exp{(-N_{\rm exp})} E \prod_{i=1}^{N_{\rm obs}}  \frac{{\rm d} \Gamma}{{\rm d} \hat{t}}(\hat{t}_{i}) \epsilon (\hat{t}_{i}) \,,
\end{equation}
where $N_{\rm obs}$ is the number of observed events and $\hat{t}_{i}$ are their timescales. PBHs are expected to have an extended mass function (see e.g. Sec. 2 of Ref.~\cite{Carr:2026hot}). Constraints calculated assuming all CO have the same mass can be translated into constraints for a specific extended mass function using the method presented in Ref.~\cite{Carr:2017jsz}.
Furthermore, as pointed out in Ref.~\cite{MACHO:1995udp}, if $f=1$ is excluded for a particular range of masses under the assumption that all CO have the same mass, then any extended mass function with $f=1$ in this range is excluded.  We note that the constraints apply to any COs in the MW halo, however (as mentioned in the introduction) astrophysical objects, such as white dwarfs, are excluded from making up a large fraction of the halo by baryon budget arguments \cite{Fields:1999ar}. Ref. \cite{Croon:2020ouk} has recalculated the microlensing constraints for extended objects, such as boson stars.

The MACHO first year results paper \cite{MACHO:1996qam} studied a very wide range of models for the MW halo, some of which are not consistent with current observations.
They found that there was a significant variation in the differential event rate, and hence the observational constraints on the CO mass and halo fraction. The OGLE collaboration finds that the uncertainty in their constraint is relatively small (see Extended Data Fig.~1 of Ref.~\cite{Mroz:2024wag}), however they only consider two benchmark models (with fixed parameter values) and assume a Maxwellian velocity distribution. Other studies \cite{Hawkins:2015uja,Green:2017qoa,Calcino:2018mwh,Garcia-Bellido:2024yaz} have found that the uncertainty in exclusion limits can be significant. Ref.~\cite{Green:2025dut} found that the rate of long-timescale events (and hence the constraints on low-mass COs) has a fairly weak dependence on the CO density and velocity distribution. However the rate of short-timescale events (and hence the constraints on high-mass COs) depends sensitively on the local DM velocity distribution. An accurate calculation of the uncertainties on the differential event rate (and parameter constraints) for models consistent with current observations, and numerical simulations of Milky Way-like galaxies, is an outstanding issue.

Microlensing observations towards the Small Magellanic Cloud (SMC) could in principle help discriminate halo-lensing from self-lensing, as the visible components of the SMC are elongated along the line of sight.
The self-lensing rate is therefore expected to be larger than for the LMC. Furthermore for halo lenses, the timescales of events towards the LMC and SMC should be similar whereas for self-lensing, events towards the SMC are expected to be longer (due to the larger lens-source separation, and hence Einstein radius) \cite{Sahu:1998rz}. There are, however, significant uncertainties in the SMC's structure due to its tidal interactions with the LMC. See Refs.~\cite{Palanque-Delabrouille:1997cxg,Graff:1998ix} for discussion of the status as of the late 1990s in the context of microlensing, and, e.g., the Introduction of Ref.~\cite{2025MNRAS.544..430P} for the current status in a more general context.

 Microlensing surveys towards the Galactic bulge
can be used to probe CO DM (see Secs.~\ref{subsec-ogle} and \ref{subsec:futureobs}). However in this case it is necessary to distinguish events due to CO DM in the MW halo from the large population of events caused by astrophysical objects in the luminous components of the MW.

Crotts explored the advantages and disadvantages of microlensing observations towards M31 (our nearest neighbour galaxy, also known as Andromeda) \cite{Crotts:1992gw}.
In particular he pointed out that the high inclination of M31's disc with respect to the line of sight would lead to a large asymmetry in the microlensing rate from lenses in the halo of M31. The event rate for the far side of the disc would be much higher than that for the near side, due to the larger column density of halo lenses. This would be a potentially powerful tool for discriminating events due to halo lenses from self-lensing or variable stars. Baillon et al. \cite{1993A&A...277....1B} pointed out that
the large number of source stars is another advantage of M31 over the MCs. However individual stars are not resolved, and hence 'pixel' lensing techniques are required (see Sec.~\ref{subsec:m31}). See Refs.~\cite{Kerins:2000xx,Baltz:2002xe} for theoretical studies of microlensing towards M31, including calculations of the differential event rate.

The standard calculation of the halo differential event rate outlined above, Eq.~(\ref{df}), assumes that the {\color{red} CO} dark matter distribution is smooth, and fully specified by the density profile, $\rho(r)$. However PBH dark matter is expected to cluster more on sub-galactic scales than particle dark matter. Since PBHs are discrete objects, there are Poisson fluctuations in their distribution and PBH clusters start forming shortly after matter-radiation equality \cite{Afshordi:2003zb,Inman:2019wvr}. For PBHs formed from gaussian fluctuations, the clusters are sufficiently diffuse \cite{Jedamzik:2020ypm} that the PBHs still act as individual microlenses. In this case the effect on the microlensing differential event rate, and hence constraints on the PBH mass and halo fraction, is small \cite{Petac:2022rio,Gorton:2022fyb}. Some PBHs are expected to be in binaries which could potentially lead to a source being lensed twice, however the binaries are typically sufficiently wide that the time between such repeating events is expected to be much longer that the duration of microlensing surveys \cite{Petac:2022rio}. The probability distribution of perturbations that are large enough to produce PBHs which make up a significant fraction of the dark matter, are expected to be non-gaussian, e.g. Ref.~\cite{Atal:2018neu}, however, and
PBHs formed from non-gaussian fluctuations can form more compact clusters \cite{Young:2019gfc}. If clusters are sufficiently compact that the cluster as a whole acts as a microlens, then microlensing parameter constraints (or allowed regions) will shift to lower PBH masses \cite{Calcino:2018mwh}.

\subsection{Non-standard events}
\label{subsec:nonstandard}

Deviations from the 'single point source, single point lens' assumptions underlying the standard Paczy\'{n}ski light curve, Eq.~(\ref{eq:amp}), can have observable consequences. 
For a more detailed review see Ref.~\cite{2012RAA....12..947M}. We discuss specific observed non-standard events in Sec.~\ref{subsec:macho}. 

The standard light curve is significantly modified if the impact parameter of an event is comparable to the ratio of the radius of the source star to the Einstein radius \cite{1994ApJ...430..505W,1994ApJ...421L..71G}.
In particular {\it finite source} effects significantly reduce the maximum magnification for COs with mass $M \lesssim 10^{-9} M_{\odot}$ \cite{Sasaki:2018dmp}.

The standard analysis assumes all motions are linear. However this is not the case if the lens or source are binary, and the Earth's orbit about the Sun also violates this assumption \cite{1992ApJ...392..442G}. The {\em parallax} effect, due to the Earth's orbit, is detectable if the event timescale is not significantly shorter than a year \cite{1992ApJ...392..442G}. For a {\it binary source}, orbital motion leads to periodic distortions in the light curve \cite{1992ApJ...397..362G}. This is commonly referred to as the 'xallarap' effect (the inverse of parallax). A {\it binary lens} can generate complex, multi-peaked light curves with caustics (where the magnification formally diverges) \cite{1991ApJ...374L..37M}.

PBHs with mass $M_{\rm PBH} \lesssim 10^{-11} M_{\odot}$ have Schwarzschild radius comparable to the wavelength of light, resulting in diffraction and interference effects. These {\it wave optics} effects reduce the magnification and place a lower limit on the mass of PBHs that can be probed by microlensing 
\cite{Sugiyama:2019dgt,Montero-Camacho:2019jte}.

Some of these effects can break the microlensing degeneracy mentioned above,
and allow the lens mass and/or projected transverse velocity to be measured. The latter is useful as it acts as a diagnostic for the lens location \cite{1996ApJ...462..705B}. For lenses in the MW halo and LMC, the projected transverse velocity is determined by the velocity dispersion of the MW halo and LMC, which are small and large respectively. For lenses in the MW disc it is determined by the differential rotation of the Sun and disk, which has an intermediate value.

\subsection{Astrometric microlensing}
\label{subsec:astrometric}
The above discussion focuses on photometric microlensing, i.e.~the change in the apparent brightness of a source during a microlensing event. This is the observable used by the microlensing surveys to date which we will overview in Sec.~\ref{sec:obs}. Astrometric microlensing is the change in the apparent position of the source \cite{1995AJ....110.1427M,1995ApJ...453...37W}, and in the future telescopes which accurately measure the positions of stars (astrometry) will also be able to probe CO DM using astrometric microlensing (see Sec.~\ref{subsec:futureobs}).

The centroid shift of the source, $\delta \theta(u)$, is given by 
\begin{equation}
\delta \theta(u) = \frac{u}{u^2 + 2} \theta_{\rm E} \,,
\end{equation}
where $\theta_{\rm E}$ is the angular Einstein radius, $\theta_{\rm E}= R_{\rm E}/D_{\rm L}$.
 Using Eq.~(\ref{re}), $\theta_{\rm E}$ can be expressed in terms of the lens-source parallax,  $\pi_{\rm rel}$,
 \begin{equation}
 \frac{\pi_{\rm rel}}{\rm mas} = \frac{\pi_{\rm L}}{\rm mas} - \frac{\pi_{\rm S}}{\rm mas} = \frac{\rm kpc}{D_{\rm L}} -  \frac{\rm kpc}{D_{\rm S}} \,,
 \end{equation}
in milliarc (mas), as
\begin{equation}
\theta_{\rm E}^2 = 8.144 \, {\rm mas} \left( \frac{M}{M_{\odot}} \right) \pi_{\rm rel} \,.
 \end{equation}
Therefore if the parallax can also be measured, the lens mass can be determined \cite{1995AJ....110.1427M,2002MNRAS.331..649B}.
Astrometric microlensing signals are detectable for large lens-source separations, so the optical depth is larger than for photometric microlensing \cite{1996ApJ...470L.113M,2000ApJ...534..213D}.

\section{Observations}
\label{sec:obs}

In this Section we first overview the general principles of stellar microlensing observations (Sec.~\ref{subsec:gen}), before looking at the results of specific surveys which have published constraints on the halo fraction of COs in Secs.~\ref{subsec:macho}-\ref{subsec:other}.
Unless otherwise mentioned, all limits (or parameter values) quoted assume that the MW halo is described by the standard halo model, see Sec.~\ref{subsec:diffeventrate}, and that all CO have the same mass. 
Sec.~\ref{subsec:summary} contains a summary of the current observational status. Figure \ref{fig-limits} shows the final published parameter constraints on the halo fraction, $f$, and CO mass, $M$. Table \ref{surveysummary} summarizes the main features of the key LMC microlensing surveys, including the number of events they observed.
This Section aims to provide theorists with an introduction to the general principles and current status of microlensing surveys. For further details see Ref.~\cite{Moniez:2010zt}, a 2010 review of microlensing from an observational perspective.

\subsection{General principles}
\label{subsec:gen}

The first step in analyzing data from a microlensing survey is to generate light curves for individual objects.
Historically this was done using point spread function photometry, e.g. Refs.~\cite{2001ApJS..136..439A,1996A&A...314...94A}. More recently different image analysis (DIA), which is more computationally demanding, but better suited to the crowded stellar fields in the MC, has been used, e.g. Refs.~\cite{Zebrun:2001gd,2000A&A...355L..39L}. 

Candidate microlensing events are selected by applying a series of cuts to the light curves. These cuts are designed to select light curves which are well fit by standard Paczy\'{n}ski form with a single, achromatic, symmetric variation as described in Sec.~\ref{subsubsec:indivdual events}, while rejecting background that can mimic microlensing. Key backgrounds are supernovae in background galaxies and variable stars \cite{MACHO:2000qbb}. A class of variable star that has recurrent bumps in their light curve that mimic microlensing, so called "blue bumpers",  were in fact discovered by microlensing surveys \cite{MACHO:1995udp}.
The detection efficiency, $\epsilon(\hat{t})$, is estimated by ascertaining what fraction of simulated events pass the selection cuts, e.g.~Ref.~\cite{2001ApJS..136..439A}. The detection efficiency is necessarily zero for timescales shorter than the interval between observations (i.e. the cadence of the survey). Typically it grows with increasing timescale to a maximum value of order $40\%$, before declining to zero again for timescales longer than the duration of the survey \cite{MACHO:2000qbb,2007A&A...469..387T,Mroz:2025xbl}.

As discussed in Ref.~\cite{Griest:2004qc}, selection criteria inevitably have to balance false positives (candidate events that are not due to microlensing) and false negatives (microlensing events that are rejected). The detection efficiency calculation takes into account the later but not the former.
Event selection (and detection efficiency estimation) is complicated by the $\sim 10\%$ of microlensing events, which are 'exotic' and don't follow the standard Paczy\'{n}ski light curve produce by a single point source and lens, e.g. Ref.~\cite{2005ApJ...631..301B}. In some cases, e.g. MACHO-LMC-9 which we discuss further in Sec.~\ref{subsec:macho}, the event is unambiguously microlensing and the features in the light curve provide information about the location of the lens. However in other cases non-microlensing variability can be fit by a non-standard lens model (see discussion in Ref.~\cite{2005ApJ...631..301B}). As discussed further in Sec.~\ref{subsec:macho}, in their final 5.7 year results, the MACHO collaboration used two sets of selection: criteria A designed to detect high-quality simple events, and criteria B designed to also select exotic and lower signal to noise events \cite{MACHO:2000qbb}. EROS-2 \cite{2007A&A...469..387T} and OGLE \cite{Mroz:2025xbl} carried out manual inspections of a subset of their light curves to check for a significant population of overlooked events. 

The measured optical depth can be estimated using
\begin{equation}
\tau = \frac{1}{E} \frac{\pi}{4} \sum_{i} \frac{\hat{t}_{i}}{\epsilon(\hat{t}_{i})} \,,
\end{equation}
where $\hat{t}_{i}$ is the estimated timescale of the $i$-th event. The estimated timescales and detection efficiency can also be used in a likelihood analysis, using Eq.~(\ref{eq:like}), to probe the CO mass and halo fraction. Note that what appears to be a single star in a survey can in fact be composed of several, unresolved, stars, a phenomenon known as blending. This affects the determination of the optical depth and event rate in multiple ways. It affects the estimation of the timescales of blended events, $\hat{t}$, and the detection efficiency, $\epsilon(\hat{t})$ \cite{Smith:2007tv}.  It also means the effective number of stars monitored, $N_{\rm \star}$, is larger than the number of catalogued objects. The impact of blending can be minimized, as is done by the EROS and OGLE collaborations, by only using bright red giant stars. See Ref.~\cite{Moniez:2010zt} for a more detailed overview.

\subsection{MACHO}
\label{subsec:macho}

Between 1992 and 1999 the MACHO collaboration monitored over 10 million stars in the LMC. Their first year results contained 3 candidate events
and (assuming all 3 were due to lens in the MW halo) they found a best-fit halo fraction $f=0.2_{-0.1}^{+0.2}$ and a MACHO mass $M= 0.065^{+0.06}_{-0.03} M_{\odot}$ \cite{MACHO:1995udp}. In their 2-year data they found 8 candidate events \cite{MACHO:1996qam}. MACHO-LMC-2 and 3 from their 1-year data set did not pass the revised event selection cuts used in the 2-year analysis (and one of these two events subsequently brightened again repeatedly and is therefore thought to be a variable star). On the other hand, 3 of the 8 2-year-candidates were in the 1-year data but did not pass the 1-year cuts \cite{MACHO:1996qam}. MACHO-LMC-9 had a U-shaped light curve, characteristic of a binary lens  
 and, if the source star is single, the lens is likely to lie in the LMC \cite{MACHO:1996xjj}. Using a sub-sample of 6 events (excluding the binary lens, MACHO-LMC-9, and MACHO-LMC-10 which they deemed a marginal candidate) they found a halo fraction $f=0.5_{-0.2}^{+0.3} $ and a MACHO mass $M=0.5_{-0.2}^{+0.3} M_{\odot}$.
A specific search for short timescale events in the 2-year data excluded COs in the range $3 \times 10^{-7} \lesssim M/M_{\odot} \lesssim 5 \times 10^{-4}$ making up all of the MW halo \cite{1996ApJ...471..774A}.
A subsequent EROS-MACHO joint analysis tightened this limit to $f < 0.10 $ for this planetary mass range \cite{1998ApJ...499L...9A}.

The final MACHO results were based on 5.7-years of observations \cite{MACHO:2000qbb}. They used two sets of selection criteria. Criteria A used similar cuts to their two year analysis \cite{MACHO:1996qam} and was designed to select simple microlensing events, with a single, high significance bump in the light curve and a flat baseline. Criteria B was intentionally looser (with higher detection efficiency), and designed to search for exotic and/or low signal-to-noise events. Marginal events suspected of being supernovae in background galaxies were included in set B but rejected from set A.
Set A contained 13 events (MACHO-LMC 1, 4, 5, 6, 7, 8, 13, 14, 15, 18, 21, 23 and 25) while an additional 4 events (9, 20, 22 and 27) passed criteria B \footnote{The event numbering reflects the fact that events 2 and 3 in the 1-year data were subsequently rejected, and a further 8 candidate events were classed as supernovae in background galaxies.}.
The best fit halo fraction (assuming a standard halo) decreased to $f \approx 0.2 \pm 0.1 $, with the best fit mass remaining of order $0.5 M_{\odot}$. The contribution of the MW disk and LMC to the total differential event rate were taken into account, with $\sim 3$ events expected to be due to stars.

Follow-up observations have subsequently revealed further information about the nature of the lenses in some events. We summarize these developments, see also Ref.~\cite{Mroz:2025aor} for more extensive, recent discussion.
Ref.~\cite{2005ApJ...631..301B} presented additional data for MACHO-LMC-4, 13 and 15. They found that all 3 events were well fit by a standard  Paczy\'{n}ski light curve, supporting their microlensing interpretation. Ref.~\cite{2005ApJ...631..301B} also discussed the interpretation of other MACHO events.
The light curve of MACHO-LMC-23 deviated from the standard Paczy\'{n}ski form, and EROS and OGLE both observed the star brightening again \cite{2005ApJ...631..301B,2007A&A...469..387T,2011MNRAS.413..493W}, so it is therefore very unlikely to be a microlensing event. 

MACHO-LMC-5 is a so called 'jerk-parallax' event \cite{Drake:2004mq}, where the light curve depends on the 2nd derivative of the Earth's velocity \cite{Gould:2003in}. Analysis of Hubble Space Telescope data allowed the parallax and proper motion of the lens to be measured, and the lens identified as an M dwarf star in the MW disk \cite{Drake:2004mq}, as suggested by Ref.~\cite{Gould:1996hv}.
High resolution follow up data from the Global Microlensing Alert Network found a periodic modulation in the light curve of MACHO-LMC-14 \cite{2001ApJ...552..259A}. These oscillations are thought to be the so-called 'xallarap' effect which occurs when the source is binary \cite{1992ApJ...397..362G}. This allowed the lens projected velocity to be estimated, and the resulting value is more consistent with expectations for a lens located in the LMC rather than the MW \cite{2001ApJ...552..259A}. The light curve of MACHO-LMC-1 has an anomaly near its peak, which suggests that the lens is binary \cite{1994AA...289L..31D,1996AA...313..841D, Mroz:2025aor}.

Follow-up infrared observations of MACHO-LMC-20 (which passed the B criteria) identified the lens as a M dwarf in the MW disk~\cite{Kallivayalil:2006yb}. Follow-up observations of MACHO-LMC-22 (which was excluded from set A by hand) found that the source is extended and has emission lines not characteristic of stars, and is therefore now thought to be a supernova or active galactic nucleus in a background galaxy \cite{Macho:2000nvd}. After MACHO-LMC-22 is excluded, none of the MACHO LMC events have Einstein diameter crossing times longer than 150 days, which excludes CO with mass $0.3 \lesssim M/M_{\odot} \lesssim 30$ making up more than $\sim 60\%$ of the MW halo \cite{Macho:2000nvd}.

Belokurov, Evans and Le Du analyzed the MACHO light curves using neutral networks and concluded that 7 of the criteria A events (MACHO-LMC-1, 5, 6, 14, 23 \footnote{Event 23 has varied again and is likely a variable star, see discussion above and Refs. \cite{2005ApJ...631..301B,2007A&A...469..387T,2011MNRAS.413..493W}} and 25) were convincing microlensing candidates \cite{Belokurov:2004am}.
Bennett also revisited the interpretation of the 13 set A MACHO events \cite{2005ApJ...633..906B}. He found that 7 of the events are confirmed microlensing candidates (MACHO-LMC-1, 4, 5, 13, 14, 15 and 25) and argued that between 10 and 12 events are likely to be microlensing (with the up to 2 remaining events being variable stars), significantly larger than the $\sim 2-3$ events expected from stars in the MW or LMC, and consistent with $f \approx 0.16 \pm 0.06$ of a standard MW halo being in CO. Subsequently, OGLE found that MACHO-LMC-7 (which was labeled 'unconfirmed' by Bennett \cite{2005ApJ...633..906B}) varied repeatedly and is hence a variable star \cite{2011MNRAS.413..493W}. OGLE also found that MACHO-LMC 8 and 18 (which were also labeled 'unconfirmed' in Ref.~\cite{2005ApJ...633..906B}) had periodic variability in their baseline \cite{Mroz:2024wag}. They argue that this is not common for microlensing events, and hence these events are likely due to variable stars, and the best fit halo fraction is in fact $\sim 40\%$ smaller than the value ($f \approx 0.2$) originally found in Ref.~\cite{MACHO:2000qbb}. Two of the MACHO set A events occurred during the operation of EROS-2 \cite{2007A&A...469..387T}.  The star in MACHO-LMC-15 was too dim to be seen by EROS-2, while MACHO-LMC-14 was observed by EROS2, but only in one colour, and was therefore not included in their sample.

As part of their response to criticism by Hawkins and Garc{\'\i}a-Bellido \cite{Hawkins:2025mlo}, the OGLE collaboration have published a reanalysis of the MACHO candidate events using more modern DIA techniques \cite{Mroz:2025aor}. They deem events MACHO-LMC-6, 13 and 15 weak candidates, due to their light curves being either asymmetric, or lacking sufficient data around the peak. They note, however, that Bennett et al.~\cite{2005ApJ...631..301B} presented additional follow-up photometry which supported the microlensing interpretation of events 13 and 15. Based on their new DIA light curves, they conclude that 7 of the MACHO events are very likely microlensing (MACHO-LMC-1, 4, 5, 9, 14, 21, 25) and argue that the other candidate events would be unlikely to pass 'sensible microlensing cuts'.

Table ~\ref{machoevents} contains a summary of the current status of the 13 events which passed criteria A. Of the 13 events, 4 are now thought to be variable stars and 2 are microlensing events with the lens likely not in the MW halo. This leaves at most 7 candidate lensing events where the lens could be located in the MW halo.

The MACHO collaboration did not publish
CO constraints from observations towards the SMC, but they did observe 2 candidate microlensing events: MACHO97-SMC-1 \cite{1997ApJ...491L..11A} (which
was also detected by EROS \cite{1998A&A...332....1P} and OGLE \cite{1997AcA....47..431U}) and MACHO98-SMC-1. The later was a caustic crossing binary lens event. Its live detection facilitated observations by multiple microlensing collaborations and an accurate measurement of a low value for the lens projected motion, indicating that the lens is likely in the SMC rather than the MW halo \cite{1999ApJ...518...44A}.

\begin{table}[H] 
\small 
\caption{A summary of the current status of the 13 events which passed criteria A in the 5.7 year MACHO collaboration analysis \cite{MACHO:2000qbb} (see also Table 1 of Ref.~\cite{Mroz:2025aor}).  The 2nd column ('Microlensing') contains information on whether the event is currently thought to be microlensing. Events classified as 'very likely microlensing' and 'weak candidate' by Ref.~\cite{Mroz:2025aor} are denoted by a green tick, \textcolor{green}{\checkmark}, and an orange question mark,  \textcolor{orange}{?}, respectively. Candidate events that are now thought to be variable stars are denoted by a red cross, \protect{\textcolor{red}{\ding{55}}}.
The 3rd column ('Lens') contains information about the lens. It is left blank where there is no information and contains 'n/a' for events which are now thought to be variable stars.} 
\label{machoevents}
\isPreprints{\centering}{} 
\begin{tabular}{
    >{\centering\arraybackslash}p{1cm}
    >{\centering\arraybackslash}p{8cm}
    c
}
\toprule
\textbf{Event}	& Microlensing & Lens	\\ 
\midrule
 1		&  \textcolor{green}{\checkmark}  &  possibly binary \cite{1994AA...289L..31D,1996AA...313..841D,Mroz:2025aor}
 \\
 4		&  \textcolor{green}{\checkmark}   & \\
 5		& \textcolor{green}{\checkmark}  & star in MW disk \cite{Drake:2004mq}\\ 
 6      &  \textcolor{orange}{?} 
 & \\ 
 7      &  \textcolor{red}{\ding{55}} variable star (has varied repeatedly) \cite{2011MNRAS.413..493W} & n/a\\
 8      &  \textcolor{red}{\ding{55}} likely a variable star \cite{Mroz:2024wag} & n/a \\
 13     &  \textcolor{orange}{?}   & \\
 14     & \textcolor{green}{\checkmark}  & likely in LMC \cite{2001ApJ...552..259A} \\ 
 15     & \textcolor{orange}{?}   & \\ 
 18     & \textcolor{red}{\ding{55}} likely a variable star \cite{Mroz:2024wag} & n/a\\ 
 21     &   \textcolor{green}{\checkmark}  & \\ 
 23     &  \textcolor{red}{\ding{55}}  brightened a 2nd time, likely a variable star \cite{2005ApJ...631..301B}  & n/a\\ 
 25      &  \textcolor{green}{\checkmark}   &  \\ 
\bottomrule
\end{tabular}

\end{table}

\subsection{EROS}

EROS ran from 1990 to 2003. Its first phase, EROS-1 found two candidate LMC events \cite{1996A&A...314...94A,1997A&A...324L..69R}. These events were compatible with microlensing by COs with mass $\approx 0.1 M_{\odot}$. However, because of the unusual nature of these events (the source star for one was an eclipsing binary star \cite{1996A&A...314...94A}), and the possibility of having detected variable stars, they chose to place an upper limits on the halo fraction: $f<0.2$ for $10^{-7} \lesssim M/M_{\odot} \lesssim 0.01$. 
A dedicated search for short-timescale events excluded planetary mass COs, with mass $10^{-7} \lesssim M/M_{\odot} \lesssim 10^{-4}$, making up more than $10\%$ of the MW halo \cite{1998A&A...329..522R}.

In their first two years of SMC observations, EROS-2 found one candidate event,  EROS2-SMC-1 \cite{1998A&A...332....1P,1999A&A...344L..63A}, also detected by MACHO (MACHO97-SMC-1) \cite{1997ApJ...491L..11A}
and OGLE \cite{1997AcA....47..431U}.
In the first 2 years of EROS-2 LMC observations they found 2 events (EROS2-LMC-3 and 4). Combined with their surviving previous 
candidate events (EROS-LMC-2 was eliminated having undergone a 2nd variation \cite{2000A&A...355L..39L}) they excluded COs with $10^{-7} (10^{-6}) \lesssim M/M_{\odot} \lesssim 1 (10^{-3}) $ making up more than $40 (10) \%$ of the MW halo \cite{2000A&A...355L..39L}.

With another three years of SMC observations EROS-2 found an additional three events (EROS2-SMC-2-4), with Einstein radius crossing times $>200$ days (longer than the events observed towards the LMC) \cite{2003A&A...400..951A}.
They argued that these events "seem more compatible with
unidentified variable stars or self-lensing within the cloud than with halo objects". This was partly based on a prior expectation that halo COs would not have super-Solar masses, however follow-up observations concluded that these events were indeed long-period variables \cite{2007A&A...469..387T}. The SMC observations placed slightly weaker constraints on the halo fraction than the previous LMC observations.

The final EROS-2 analysis used observations of a subsample of 7 million bright stars in the MC, in order to aid discrimination of variable stars, and minimize the effects of blending \cite{2007A&A...469..387T}. They found only one event, and constrained COs with $10^{-6} \lesssim M/M_{\odot} \lesssim 1$ to make up less than $10\%$ of the MW halo. Section 6 and Table 4 of Ref.~\cite{2007A&A...469..387T} contains an update on the status of the 11 previously published EROS MC events. Only 1 (EROS-SMC-1) remains a microlensing candidate. EROS1-LMC-1 and 2, and EROS2-LMC-3 and 4 underwent a 2nd variation. With improved data analysis EROS2-LMC-5, 6 and 7 were classified as Supernovae, and (as mentioned above) additional data showed that EROS2-SMC-2, 3 and 4 were long period variable stars. 

Ref.~\cite{Blaineau:2022nhy} combined archival data from EROS-2 and MACHO to probe long timescale events, and concluded that CO with mass $10 \lesssim M/M_{\odot} \lesssim 100 (10^{3})$ have $f< 0.15 (0.5)$. Combined with the previous EROS results, $f>0.15$ was excluded for  $10^{-6} \lesssim M/M_{\odot} \lesssim 100$.

\subsection{OGLE}
\label{subsec-ogle}
The OGLE project has run for more than thirty year, starting in 1992 \cite{2024CoSka..54b.234S}.
The first OGLE MC observations (OGLE-II) found two candidate LMC events \cite{Wyrzykowski:2009ep}. They found that if both were due to COs the halo fraction was $f = 0.08 \pm 0.06$. However they argued that these events were consistent with expectations from self-lensing, and constrained CO with $0.01 \lesssim M/M_{\odot} \lesssim 0.2$ to make up less than 10\% of the MW halo. OGLE-LMC-02 subsequently underwent multiple additional brightenings and is therefore not a microlensing event \cite{Mroz:2024wag}.
Their initial observations towards the SMC found a single, unconvincing, candidate and slightly weaker limits \cite{Wyrzykowski:2010bh}.

In 8 years of observations towards the SMC, OGLE-III found 3 convincing events (OGLE-SMC-2-4) \cite{Wyrzykowski:2011tr}. The light curve of OGLE-SMC-2 exhibits a clear parallax effect, and also deviates from the simple Paczy\'{n}ski curve around its peak. Follow up observations (including data from the Spitzer space telescope) measured the lens projected velocity and found that the lens is most likely a BH binary in the MW halo, with mass $M \sim 10 M_{\odot}$ \cite{Dong:2007px}. Ref.~\cite{Wyrzykowski:2011tr} argued that this event is consistent with BHs making up of order a per-cent of the MW halo. Like MACHO-LMC-5, OGLE-SMC-3 was a jerk-parallax event, with the lens likely a M dwarf in the MW disc \cite{Wyrzykowski:2011tr}. The combined OGLE-II and III MC limits on the halo fraction were similar to those from EROS2 for $M \gtrsim 0.01 M_{\odot}$ \cite{Wyrzykowski:2011tr}.

Results from OGLE-III and IV observations of 78.7 million stars in the LMC for up to 20 years were published in 2024 \cite{Mroz:2024wag}. Their automated data pipeline found 13 events, all with Einstein radius crossing times less than one year. A manual, 'by eye', search found an additional 3 events.
They found that the number of events, and their positions, timescales and parallaxes, are consistent with expectations from stars in the MW disk and the LMC. Using the 13 events identified using the automated data pipeline, they found that COs in the mass range $ 10^{-4} (10^{-5}) \lesssim M/M_{\odot} \lesssim 6 (900) $ can not make up more than $1 (10)\%$ of the MW halo (assuming the MW halo has a contracted Navarro Frenk White density profile fitted to Gaia DR2 and other data in Ref.~\cite{Cautun:2019eaf}) \cite{Mroz:2024mse}. There are also similar, but slightly weaker, constraints from OGLE's 20 years of SMC observations, with the six observed events being consistent with expectations from stars in the SMC or MW disc \cite{Mroz:2025xbl}.

Niikura et al.~\cite{Niikura:2019kqi} used OGLE microlensing observations towards the Galactic bulge to probe planetary mass COs. They found that most events were well-fit by microlensing by stellar populations, and excluded $f>0.1$ for $10^{-6} \lesssim M/M_{\odot} \lesssim 10^{-2}$. 
There was a distinct population of six ultra-short ($\sim 0.1$ day) events. These events could be due to free-floating planets, however they argued that these events are also consistent with Earth mass COs making up a few per-cent of the MW halo. From 2022-2024 OGLE carried out a high-cadence survey of the MC, observing 35 million stars up to 21 times per night \cite{Mroz:2024wia}. They only found one event, with Einstein radius crossing time $\sim 90$ days, which allowed them to exclude COs with mass $10^{-8} \lesssim M/M_{\odot} \lesssim 0.01$ making up more than $1\%$ of the MW halo.

\subsection{Surveys focused on M31}
\label{subsec:m31}

Various collaborations have carried out so-called pixel microlensing surveys towards M31. Stars in M31 are too distant to be resolved individually, so instead microlensing surveys search for variations in the flux of the pixels of the image. Ref. \cite{CalchiNovati:2009af} contains a review of pixel microlensing surveys towards M31 as of 2009.

In the early 2000s POINT-AGAPE \cite{POINT-AGAPE:2005swi} found 6 events, significantly more than the $\sim 1$ they expected from self-lensing. They concluded that if the average mass of COs is $\sim  1 (0.01) M_{\odot}$, then at 95\% confidence they make up more than $20 (8) \% $ of the MW halo. Subsequently, MEGA \cite{2006A&A...446..855D} and PLAN \cite{2014ApJ...783...86C} 
both concluded (using detailed calculations of the detection efficiency, and also more sophisticated mass models of M31 to calculate the self-lensing rate) that their observed event rates were broadly consistent with self lensing, and constrained the MW halo fraction in Solar mass COs to be less than $(20-30) \%$.

More recently high cadence (2 minute sampling interval) observations of M31 have been carried out using the Hyper Suprime-Cam (HSC) on the Subaru telescope \cite{Niikura:2017zjd,Sugiyama:2026kpv}. Initial observations found one candidate event which led to a constraint $f \lesssim 10^{-2}$ for $10^{-9} \lesssim M/M_{\odot} \lesssim 10^{-5}$ \cite{Niikura:2017zjd,Niikura:2017zjd}. The finite size of the lensed stars reduces the magnification \cite{1994ApJ...430..505W}, and makes the constraint weaker than it would otherwise be, for $M \lesssim 10^{-7} M_{\odot}$ \cite{Niikura:2017zjd,Smyth:2019whb}. Very recently this work has been updated with additional data, and improved analysis of the original data (in particular the inclusion of finite source size effects in the detection efficiency estimation) \cite{Sugiyama:2026kpv}.
Twelve microlensing candidates have been identified, with 4 being classed as secure, high-significance, detections. If these events are due to COs in the MW halo the corresponding mass and halo fraction are $M \sim (10^{-7} - 10^{-6}) M_{\odot}$ and $f \sim 0.01-0.1$. Mr\'{o}z and Udalski have, however, carried out an independent reanalysis of the data and argue that all 12 candidate events are in fact variable stars \cite{Mroz:2026nez}.  
\\

\subsection{Kepler}
\label{subsec:other}
In the early 2010s Griest and collaborations searched for short timescale microlensing events in publicly available observations of nearby stars (designed to search for extra-Solar planets) by Kepler. Their final results~\cite{Griest:2013aaa}, based on 2 years of data, excluded compact objects with $10^{-9} \lesssim M/M_{\odot} \lesssim 10^{-7}$ making up all of the MW halo. 

\subsection{Summary}
\label{subsec:summary}

\begin{figure}[htbp]
\isPreprints{\centering}{} 
\includegraphics[width=10 cm]{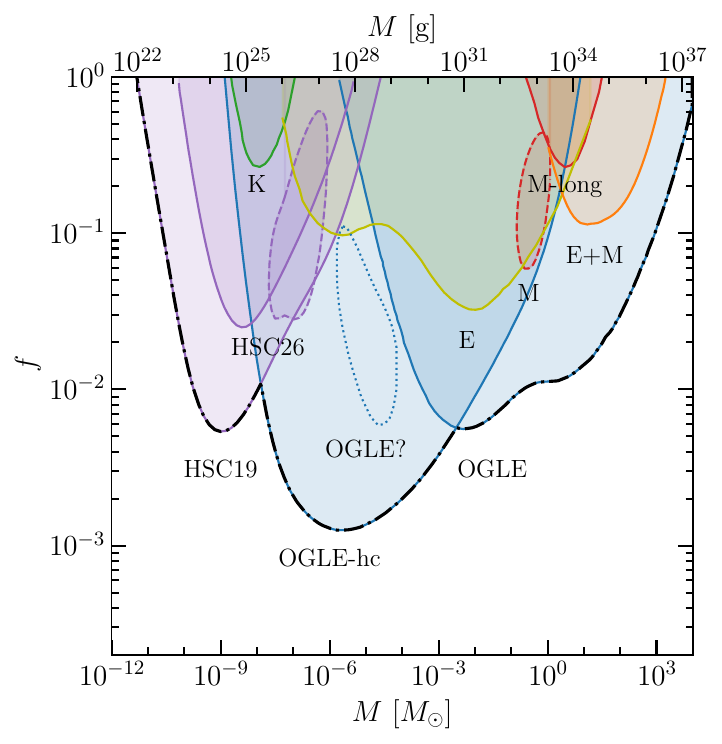}
\caption{Constraints on the CO mass, $M$, and MW halo fraction, $f$, at 95\% confidence, assuming all CO have the same mass. The solid lines denote exclusion limits, while the dot-dashed black line shows the envelope of the tightest constraint for each value of $M$. The dashed lines show allowed regions of parameter space.
The exclusion limits are from EROS ('E') \cite{2007A&A...469..387T} (yellow), MACHO long timescale ('M-long') \cite{Macho:2000nvd} (red), EROS plus MACHO long timescale ('E+M') \cite{Blaineau:2022nhy} (orange),  OGLE \cite{Mroz:2024wag} and OGLE high cadence ('OGLE-hc') \cite{Mroz:2024wia} (blue), and  
Kepler ('K') \cite{Griest:2013aaa} (green). 
The purple lines are from high cadence observations of M31 using the Subaru-HSC. 'HSC19' shows the original exclusion limit \cite{Niikura:2017zjd} as calculated in Ref.~\cite{Croon:2020ouk}, 
while 'HSC26' shows the exclusion limit from the latest analysis, under the assumption that none of the events observed are due to CO in the MW halo. The dashed line shows the allowed parameter values assuming that the 4 'secure' high-significance events are due to CO in the MW halo. The dashed red line shows the allowed parameter values from the 13 set A events in the final 5.7 year MACHO data set, as published in Ref.~\cite{MACHO:2000qbb}. Subsequent observations indicate that some of these events are not microlensing (see Sec.~\ref{subsec:macho} and Refs.~\cite{2007A&A...469..387T,Mroz:2025aor} for discussion). 
The dotted blue line, labeled 'OGLE?', shows the parameter values found assuming that the ultra-short duration events observed towards the Galactic bulge by OGLE are due to COs in the MW halo~\cite{Niikura:2019kqi}. 
This plot was made using a modified version of Kavanagh's PBHBounds \cite{PBHbounds}. \label{fig-limits}}
\end{figure}   

Figure \ref{fig-limits} shows the final published parameter constraints on the halo fraction, $f$, and compact object mass, $M$, (assuming all compact objects have the same mass) from the surveys discussed in this section \footnote{Results from M31 surveys, other than HSC, are not shown. These surveys quoted approximate allowed or excluded values for the halo fraction, but they did not publish exclusion limit curves or allowed regions of parameter space.}. The solid lines denote exclusion limits, while the dot-dashed black line shows the envelope of the tightest constraint for each value of $M$. The dashed lines show allowed regions of parameter space.
Table \ref{surveysummary} contains a summary of the key features of the MACHO \cite{MACHO:2000qbb}, EROS \cite{2007A&A...469..387T} and OGLE \cite{Mroz:2024wag} LMC microlensing surveys. See Table 1 of Ref.~\cite{Moniez:2010zt} for a more detailed comparison of stellar microlensing surveys (as of 2010).

\begin{table}[H] 
\small 
\caption{A summary of the main features of the MACHO \cite{MACHO:2000qbb}, EROS \cite{2007A&A...469..387T} and OGLE \cite{Mroz:2024wag} LMC microlensing surveys. The 13 MACHO events are from their tighter selection criteria A. Four of these events have subsequently been found to likely be variable stars, see Ref.~\cite{2007A&A...469..387T,Mroz:2025aor} and Sec.~\ref{subsec:macho}. The 13 events listed for OGLE are those found using their automated search pipeline, and used for their halo fraction constraints, see Sec.~\ref{subsec-ogle}.  \label{surveysummary}}
\isPreprints{\centering}{} 
\begin{tabularx}{\textwidth}{CCCCCC}
\toprule
\textbf{Survey}	& Duration  & Field of view & Number of stars, $N_{\star}$ & Number of events, $N_{\rm obs}$	\\ 
& (years) & (${\rm deg}^2$) & (million)  \\
\midrule
MACHO & 5.7 & 14 & 12 & 13 \\
EROS & $\sim$ 12 & 88 (EROS-2) & $\sim$ 25 (EROS-2)& 1\\
OGLE & $\sim$ 20 & 105 (OGLE-IV)& 78.7 (OGLE-IV) & 13\\
\midrule
\bottomrule
\end{tabularx}
\end{table}

\section{Future}
\label{sec:future}

\subsection{Observations}
\label{subsec:futureobs}

In the near future new telescopes will be able to carry out next generation microlensing surveys. The Nancy Grace Roman Space Telescope (formerly known as WFIRST) has excellent resolution, which will reduce the effects of blending and allow the measurement of finite source effects. It will carry out high cadence observations, as required to detect short timescale events and probe light CO. The ground-based Vera C. Rubin Observatory (formerly known as the Large Synoptic Survey Telescope) will carry out a long duration wide-field survey. It will have high sensitivity to long timescale events, and hence intermediate and high mass CO. 
Telescopes with high spatial resolution, capable of high-precision astrometry, will be able to carry out astrometric microlensing measurements and probe the angular Einstein radius (see Sec.~\ref{subsec:astrometric}).

To probe CO DM with  
the Roman Space Telescope's Galactic Bulge Time Domain Survey it will be necessary to discriminate events due to CO in the MW halo from events due to astrophysical objects in the luminous components of the MW.
Following early work in Ref.~\cite{2013ApJ...767..145C}, Ref.~\cite{DeRocco:2023gde} showed that it will be possible to discriminate planetary-mass CO DM from free-floating planets, and probe halo fractions $f\sim 10^{-3}-10^{-4}$ using just the timescales of (photometric) microlensing events. 
Ref.~\cite{Fardeen:2023euf} showed that this survey could probe  
$f\sim 10^{-3}$ for $10^{-1} \lesssim M/M_{\odot} \lesssim 10^{2}$ {\it if} it is possible to distinguish CO DM and stellar lenses using the joint distributions of the timescales and angular Einstein radii of events. If it is not possible to do this then the sensitivity will be reduced to $f\sim 10^{-1}$, weaker than existing OGLE limits in this mass range \cite{Mroz:2024wag}. 
Similar constraints may also be possible using the Rubin Observatory \cite{LSSTDarkMatterGroup:2019mwo}.
With improved understanding of backgrounds, Gaia time-series data could probe halo fractions down to $f\sim 10^{-3}$ for multi-Solar mass CO DM \cite{Verma:2022pym}, see also Ref.~\cite{2002MNRAS.331..649B}. Thirty-metre class, Extremely Large Telescopes, will also be able to probe CO DM via astrometric microlensing \cite{Bird:2022wvk}.
Combining photometric and astrometric microlensing data would aid the detection and characterization of events \cite{Pruett:2022ber}. In all cases, correctly identifying events and minimizing the false positive rate will be crucial (as we have seen for previous surveys in Sec.~\ref{sec:obs}). See Ref.~\cite{CrispimRomao:2025kxx} for work in this direction.

\subsection{Theory}
\label{subsec:futuretheory}
As discussed in Sec.~\ref{subsec:diffeventrate}, uncertainties in the DM distribution lead to uncertainties in the differential event rate and hence the constraints on the halo fraction.
If the DM is smoothly distributed, the most significant uncertainty is the dependence of the rate of short-timescale events, and hence the constraints on $f$ for large masses, on the local velocity distribution \cite{Green:2025dut}. In the case of PBHs, if a significant fraction are in compact clusters, such that the cluster as a whole lenses, the constraints would change significantly \cite{Calcino:2018mwh}. Understanding the present day spatial distribution of PBHs formed from realistic density perturbation distributions (i.e. how many PBHs are in clusters, and what are their properties? how many PBHs are in binaries, and what are their properties?) is a major outstanding challenge.

\section{Summary}
\label{sec:summary}

We have explored how stellar microlensing surveys probe compact object dark matter, in particular primordial black holes. Early indications from the MACHO collaboration that a significant fraction of the Milky Way halo could be in the form of roughly Solar Mass COs \cite{MACHO:1996qam,MACHO:2000qbb} were not confirmed by EROS \cite{2007A&A...469..387T} and OGLE \cite{Mroz:2024wag}. Long duration \cite{Mroz:2024wag} and high cadence \cite{Niikura:2017zjd,Mroz:2024wia} observations have extended the range of CO masses probed by microlensing. Under the standard assumptions outlined in Sec.~\ref{sec:theory}, collectively these surveys exclude COs with mass $10^{-11} \lesssim M/M_{\odot} \lesssim 10^{4}$ making up all of the DM in the MW, while those in the mass range $10^{-10} \lesssim M/M_{\odot} \lesssim 1$ can make up less than one percent. Further theoretical work is required to ascertain how robust the standard assumptions regarding the CO distribution are. For PBHs a key question is whether a significant fraction are in compact clusters. Future observations, in particular astrometric ones using the Roman Telescope and Rubin Observatory, can potentially probe CO dark matter down to $f \sim 10^{-3}-10^{-4}$. However further work is required to develop techniques for discriminating CO dark matter-induced lensing events from backgrounds, including lensing events due to known astrophysical objects, such as stars and planets.





\funding{This research was supported by an STFC Consolidated Grant [Grant No. ST/T000732/1] and a Leverhulme Trust Research Fellowship [Grant No. RF-2025-282]. 
}

\dataavailability{No new data were created or analyzed in this study. Data sharing is not applicable to this article.}


\conflictsofinterest{The author declares no conflicts of interest.} 


\abbreviations{Abbreviations}{
The following abbreviations are used in this manuscript:
\\

\noindent 
\begin{tabular}{@{}ll}
CO & Compact Object \\
DM & Dark Matter \\
DIA & Difference Image Analysis \\
LMC & Large Magellanic Cloud \\
MC & Magellanic Cloud\\
MW & Milky Way \\
PBH & Primordial Black Hole\\
SMC & Small Magellanic Cloud
\end{tabular}
}

\appendixtitles{no} 



\reftitle{References}


\bibliography{PBHmicro}


\PublishersNote{}
\end{document}